\documentclass[aps,prl,tightenlines,twocolumn,superscriptaddress,showpacs]{revtex4}

\usepackage{amssymb}
\usepackage{color,graphicx}
\usepackage{amsmath}
\usepackage{amsbsy}
\usepackage{amsthm}
\usepackage{bbm}
\usepackage{bm}
\usepackage{epsfig}

\newcommand{\be}{\begin{equation}}
\newcommand{\ee}{\end{equation}}
\newcommand{\beq}{\begin{eqnarray}}
\newcommand{\eeq}{\end{eqnarray}}

\begin{document}

\title{On the uniqueness of the equation for state-vector collapse}

\author{Angelo Bassi}
\email{bassi@ts.infn.it} \affiliation{Department of Physics, University of
Trieste, Strada Costiera 11, 34151 Trieste, Italy} \affiliation{Istituto
Nazionale di Fisica Nucleare, Trieste Section, Via Valerio 2, 34127 Trieste,
Italy}

\author{Detlef D\"urr}
\email{duerr@mathematik.uni-muenchen.de}
\address{Mathematisches Institut der L.M.U., Theresienstr. 39, 80333
M\"unchen, Germany.}

\author{G\"unter Hinrichs}
\email{hinrichs@math.lmu.de}
\address{Mathematisches Institut der L.M.U., Theresienstr. 39, 80333
M\"unchen, Germany.}

\begin{abstract}
The linearity of quantum mechanics leads,  under the assumption that the wave function offers a complete description of reality, to grotesque situations famously known as Schr\"odinger's cat. Ways out are either adding elements of reality or replacing the linear
evolution by a nonlinear one.  Models of spontaneous wave function collapses took the latter path.  
The way such models are constructed leaves the question, whether such models are in some sense unique, 
i.e. whether the nonlinear equations replacing Schr\"odinger's equation, are uniquely determined as collapse equations. 
Various people worked on identifying the class of nonlinear modifications of the Schr\"odinger equation, compatible with general physical requirements. 
Here we identify the most general class of continuous wavefunction evolutions under the assumption of no-faster-than-light signalling.
\end{abstract}
\maketitle

\emph{Introduction.}
The superposition principle of quantum mechanics has always been an open issue. From the theoretical point of view, the strongest reason is that the collapse of the wave function, which has been formulated to avoid macroscopic superpositions, is clearly an artifact~\cite{bell}. To overcome this situation, collapse models have been devised, which incorporate wave-function's collapse, together with the Schr\"odinger evolution, into one single (nonlinear and stochastic) dynamical law~\cite{grw,csl,diosi,physrept,sc1,rmp,Weinberg}. Independently from this, Weinberg questioned whether quantum mechanics, being a linear theory, should be regarded as an approximation of an underlying non-linear theory~\cite{weinl}, pretty much in the same way in which Newtonian gravity is a weak-limit limit of general relativity. Moreover when one combines quantum mechanics and gravity even in the most naive way, by taking into account the gravitational interaction of the wave function with itself, one obtains  nonlinear evolutions~\cite{sn1,sn2,sn3}. 

Quantum linearity has been challenged also experimentally, and the interest and efforts are constantly increasing. Perhaps the most famous tests are diffraction experiments with macro-molecules~\cite{arndt1,arndt2,arndt3}. In the most recent experiment of this kind, quantum coherence was proven to hold for molecules with more than 10,000 amu~\cite{arndt3}. The question remains open whether larger systems still enjoy this property. Other proposals, representing a direct test of quantum linearity, include satellite atom (Cs)~\cite{exp1}, micro-mirrors~\cite{exp2}, nano-sphere interferometric experiments~\cite{exp3}. To these, one should add the vast zoo of indirect tests of the superposition principle~\cite{lub,sc1}.

Among all proposals for nonlinear modifications of quantum mechanics, only collapse models survived. The most important reason is that, in response to Weinberg's attempt to modifying quantum theory in a nonlinear sense, N. Gisin~\cite{gisin,gisin2}, followed by J. Polcinski~\cite{pol}, proved that any nonlinear and deterministic modification of the theory yields superluminal signalling. Therefore only {\it stochastic} nonlinear modifications are allowed. Collapse models fall in this class. However, it is not clear at all whether they should be the only possible models, since their nonlinear structure is very specific and  given by the following stochastic differential equation~\cite{physrept}:
\begin{eqnarray} \label{eq:sdgr}
\text{d} \psi_t & =  & \left[ -i H \text{d}t + \sum_{k=1}^{n} \left(  L_k - \ell_{k,t} \right)\text{d} W_{k,t} \right. \nonumber \\
& & - \left. \frac{1}{2} \sum_{k=1}^{n} \left(L^{\dagger}_k L^{\phantom{\dagger}}_k  - 2 \ell_{k,t}L_k + |\ell_{k,t}|^2\right)\text{d}t\right] \psi_t, \quad \\
& & \nonumber \\
\ell_{k,t} & \equiv & \displaystyle \frac12\langle\psi_t,(L^{\dagger}_k + L^{\phantom{\dagger}}_k)\psi_t\rangle
\end{eqnarray}
where $H$ is the standard quantum  Hamiltonian of the system, $L_k$ are linear operators defining the preferred basis for the collapse, and $W_{k,t}$ standard Wiener processes. As one can see, nonlinearity enters in a very specific way. Of course the above equation can be generalized, e.g. by by considering other kinds of noise. One can also consider jump-type models, like the GRW model~\cite{grw}, instead of diffusion processes. Nevertheless, the basic structure does not change. 

In this paper we classify all Markovian wave function evolutions~\cite{comment} which do not allow faster-than-light signalling. Our analysis is naturally split into two steps. In the first step we show that, under  reasonable conditions on the evolution of the statistical operator (emerging from the no-faster-than-light-signalling condition),
the latter evolution is of Lindblad form. A remarkable fact is that complete positivity is not required, but comes as a byproduct. 
In the second step, we characterize the wave function diffusions which lead to a Lindblad-type of evolution for the statistical operator. These diffusions practically coincide with those of collapse models.

Similar question have been dealt with before in the literature: 
S.L. Adler and T.A. Brun were the first to raise the question whether indeed collapse models are the only models compatible with the no-faster-than-light 
requirement~\cite{adlernftlr,adlernftlr2}.
Given the Lindblad form for the evolution of the statistical operator, the diffusive unravelings have been considered
 by Gisin and Percival \cite{pergi}, who used only complex Wiener processes and hence obtained only a partial answer.
In particular, the most popular collapse models are not included. 
Adler and Brun also provided a partial answer to that question, their proof being restricted to the very special case of linear operators~\cite{op} in the 
stochastic differential equation for the wave function.
H.M. Wiseman and L. Diosi ~\cite{WisemanDiosi} answered that question in the same generality as we do. For completeness of the presentation we give a 
proof  in our setting in the appendix. Their result looks different from ours, but under noise transformations the results come out the same.
Recently, also S. Weinberg~\cite{Weinberg} raised the same question and given the renewed interest in the subject. We think it is helpful if now an 
exhaustive answer is given. 
We shall therefore prove:

\noindent \textit{Theorem 1:
 Assume that a random (normalized) wave function $\psi_t$ in the finite-dimensional Hilbert space $\mathbb C^d$ evolves in a Markovian way and satisfies two further conditions:
\begin{enumerate}
 \item The evolution of $P_t:=|\psi_t\rangle\langle\psi_t|$ is Markovian.
 \item The evolution of the statistical operator $\rho_t:=\mathbb E|\psi_t\rangle\langle\psi_t|$ ($\mathbb E$ denotes the stochastic average) is closed and linear in the sense that, 
whenever $s<t$ and
$\sum_i\lambda_i|\chi_i\rangle\langle\chi_i| = \sum_i\mu_i|\phi_i\rangle\langle\phi_i| =: \rho$ ,
\begin{align*}
 &\sum_i\lambda_i\mathbb E\left(|\psi_t\rangle\langle\psi_t| \mid \psi_s=\chi_i\right) \\
=& \sum_i\mu_i\mathbb E\left(|\psi_t\rangle\langle\psi_t| \mid \psi_s=\phi_i\right)
\end{align*}
holds, which implies that the propagator
\begin{align*}
\mathcal U_{t,s}\rho:=& \sum_i\lambda_i\mathbb E\left(|\psi_t\rangle\langle\psi_t| \mid \psi_s=\chi_i\right) \\
=& \sum_i\lambda_i\mathbb E\left(P_t \mid P_s=|\chi_i\rangle\langle\chi_i|\right)
\end{align*}
is a well-defined linear operator.
\end{enumerate}
Then $\mathcal U_{t,s}$ forms a time-homogenous completely positive quantum-dynamical semigroup, i.e. $\rho_t$ satisfies a Lindblad equation~\cite{Lin}:
\begin{equation}\label{Lindblad2}
 \frac{\text d\rho_t}{\text dt} = -i[H,\rho_t] + \sum_{k=1}^n\left(L_k\rho_tL_k^{\dagger} - \frac12 L_k^{\dagger}L_k\rho_t - \frac12 \rho_tL_k^{\dagger}L_k\right)\!.
\end{equation}
}

\noindent \textit{Remark:} Since we consider at this stage general nonlinear evolutions of the wave functions it could happen that the global phase of the wave function determines the evolution of the wave function. 
Since in the operator  $P_t$  the initial global phase drops out, the evolution of $P_t$ need not be Markovian anymore. Therefore we need to require Property 1.
Property 2 arises from Gisin's theorem~\cite{gisin,gisin2}, which states that, whatever the evolution for the state vector,
the equivalence among statistical ensembles~\cite{equiv} must be preserved in order to avoid faster-than-light signalling. This implies a closed and linear evolution of the statistical operator.
Note that complete positivity is not part of the assumptions. It comes as a necessity from the very fact that we have assumed a (Markovian) dynamics for the wave function.

Before giving the proof of the theorem, we turn now to the second part of the problem.
\color{black}


\noindent {\it Question:} {\it Given a diffusion process for the wave-function:
\begin{equation}\label{collapse2}
 \text d\psi_t=A(\psi_t)\text dt + \sum_{k=1}^NB_k(\psi_t)\text dW_{k,t},
\end{equation}
where $A(\psi)$ and $B_k(\psi)$ are unspecified nonlinear operators, find the conditions under which it generates a Lindblad type of equation
for the statistical operator $\rho_t \equiv {\mathbb E}[|\psi_t \rangle\langle \psi_t|]$.}


 We shall answer the question by proving the following result.

\noindent {\it  Theorem 2: Assume that $1, L_1, \dots, L_n$ are linearly independent. Then  Eq.~\eqref{collapse2} leads to Eq.~\eqref{Lindblad2} if and only if $N\ge n$ and the operators $A(\psi)$ and $B_k(\psi)$ take the following form:
\begin{eqnarray}
A(\psi) & = & \displaystyle -iH -\frac{1}{2}\sum_{k=1}^{N} \left(   L^{\dagger}_k L^{\phantom{\dagger}}_k  - 2 \ell^{(\psi)}_{k,t}L^{(\psi)}_k + |\ell^{(\psi)}_{k,t}|^2\ \right)      \;\;\;\;          \\
& & \nonumber \\
B_k(\psi) & = & L^{(\psi)}_k - \ell^{(\psi)}_{k,t}, \\
& & \nonumber \\ 
\ell^{(\psi)}_{k,t}  & = &  \displaystyle \frac12\langle\psi_t,(L^{(\psi)\dagger}_k + L^{(\psi)}_k)\psi_t\rangle 
\end{eqnarray} 
modulo unimportant global phase factors. Here $L_{n+1},\dots, L_N:=0$ and the operators $L^{(\psi)}_k$ and their ``adjoints'' are defined as follows,
\begin{equation}\label{eq:fgdf}
L^{(\psi)}_k \; := \; \sum_{j=1}^{N} \, u_{kj}(\psi)\, L_j, \qquad L^{(\psi)\dagger}_k \; := \; \sum_{j=1}^{N} \, u_{kj}^*(\psi)\, L_j^{\dagger}
\end{equation}
where $u_{kj}(\psi)$ are the unknown coefficients of a $N\times N$ unitary matrix. This unitary freedom, which is a symmetry of the Lindblad equation~\eqref{Lindblad2}, implies that  the allowed diffusions range from linear diffusion (no collapse) to the usual collapse equation. The collapse rate for the diffusion is maximal for the usual collapse equation~\eqref{eq:sdgr}.}


\noindent Theorem 1 together with theorem 2 is a remarkable result: 
It essentially says that the usual collapse models are the only possible nonlinear (Markovian) extensions of the Schr\"odinger equation, 
compatible with the no-faster-than-light assumption.

Note that the above result is completely general. It includes unravellings via $n$ real Wiener processes and, setting $N=2n$, those via $n$ complex ones as well.
For example, if $n=1$ and $L^\dagger=L$, then $u_{11}=1$ gives the standard collapse unravelling whereas, adding $L_2=0$ and using $u=\frac1{\sqrt2}\begin{pmatrix}1&-i\\i&1\end{pmatrix}$, one obtains:
\begin{eqnarray}\label{diosicomplex}
\text{d} \psi_t  & = &  \left[ -i H \text{d}t + \left( L - \ell_{t} \right)\text{d} \frac{W_{1,t}+iW_{2,t}}{\sqrt2}  \right. \nonumber \\
& & \left. - \frac{1}{2} \left(L - \ell_{t}\right)^2\text{d}t\right] \psi_t
\end{eqnarray}
with independent real Wiener processes $W_{t,1}$ and $W_{t,2}$. Even unravellings making use of correlated Wiener processes $W_{t,1}$ and $W_{t,2}$ are included because they can be obtained by a linear transformation from uncorrelated ones and this linear transformation can be absorbed in the $B_k$ terms.
Eq.~\eqref{diosicomplex} is invariant under the symmetry transformations of the Lindblad equations as opposed to the standard unravelling.
When one considers the state vector as fundamental element of the theory, as we do,  the symmetries of the Lindblad equation need not be obeyed by the equation for the state vector.  For us the latter defines the relevant symmetries, while in~\cite{WisemanDiosi} the symmetries of the Lindblad equation seem to be considered as fundamental.

\emph{Proof of theorem 1.}
The time-homogeneity of $\psi_t$ is immediately carried over to $\mathcal U_{t,s}$. Write $\mathcal U_t:=\mathcal U_{t,0}$, chose a statistical operator $\rho$ and find a representation
$\rho=\sum_i\lambda_i|\chi_i\rangle\langle\chi_i|$. Then
\begin{equation*}
 \begin{split}
  \mathcal U_t\mathcal U_s\rho =& \; \mathcal U_t\sum_i\lambda_i\mathbb E\left(P_s\mid P_0=|\chi_i\rangle\langle\chi_i|\right) \\
=& \sum_i\lambda_i\mathbb E\left(\mathcal U_tP_s\mid P_0=|\chi_i\rangle\langle\chi_i|\right) \\
=& \sum_i\lambda_i\mathbb E\left( \mathbb E(P_{t+s}\mid P_s) \mid P_0=|\chi_i\rangle\langle\chi_i|\right) \\
=& \sum_i\lambda_i\mathbb E\left( \mathbb E(P_{t+s}\mid P_s, P_0=|\chi_i\rangle\langle\chi_i|) \mid P_0=|\chi_i\rangle\langle\chi_i|\right) \\
=& \sum_i\lambda_i\mathbb E\left( P_{t+s} \mid P_0=|\chi_i\rangle\langle\chi_i|\right) \\
=& \; \mathcal U_{t+s}\rho \,,
 \end{split}
\end{equation*}
which is the semigroup property.

Since $\mathcal U_t$ is linear and (obviously) trace-preserving, $\rho_t$ satisfies
\begin{equation*}
 \frac{\text d}{\text dt}\rho_t = -i[H,\rho_t] + \!\!\sum_{i,j=1}^{d^2-1}\!\! c_{ij} \! \left(F_i\rho_tF_j^{\dagger} \! - \! \frac12 F_j^{\dagger}F_i\rho_t \! - \! \frac12 \rho_tF_j^{\dagger}F_i\right)
\end{equation*}
with a trace-free hermitian operator $H$, operators $F_j$ such that $1, F_1, F_2, \dots$ are linearly independent and a hermitian matrix $(c_{ij})$ (see \cite{GKS}).
Diagonalizing $(c_{ij})$ via a unitary matrix $(u_{ij})$, so that $c_{ij}=\sum_ku_{ik}c_ku_{jk}$, and defining $L_k:=\sum_i u_{ik}L_i$, so that also $1,L_1, L_2, \dots$ are linearly independent,
one obtains
\begin{equation}\label{lindblad+-}
 \frac{\text d}{\text dt}\rho_t = -i[H,\rho_t] + \!\! \sum_{k=1}^{d^2-1} \!\! c_{k} \! \left(L_k\rho_tL_k^{\dagger} \! - \! \frac12 L_k^{\dagger}L_k\rho_t \! - \! \frac12 \rho_tL_k^{\dagger}L_k\right)
\end{equation}
with real constants $c_k$.

For the rest of the proof, we fix $\varphi\in\mathbb C^d$ such that $\varphi, L_1\varphi, \dots$ are linearly independent and an orthonormal basis $(e_k)$ of $\mathbb C^d$ with $e_1=\varphi$.
Substituting $\rho_t=\mathcal U_t|\varphi\rangle\langle\varphi|$ and denoting restrictions of vectors and operators to the orthogonal complement of $\varphi$ by $\bot$, Eq.~\eqref{lindblad+-} at $t=0$ gives
\begin{equation}\label{senkrecht1}
\left. \frac{\text d}{\text dt}\right|_{t=0}\rho_{t,\bot} = \sum_k c_k|L_k\varphi\rangle_\bot\langle L_k\varphi|_\bot \,.
\end{equation}

Another restriction on the evolution of $\rho_t$ comes from the Markov property of $\psi_t$: According to Courr\`ege's theorem, for any smooth function $f$, the function $f_t(\psi):=\mathbb E(f(\psi_t)\mid\psi_0=\psi)$
evolves according to:
\begin{equation}\label{Courrege1}
\frac{\partial f}{\partial t}=Lf
\end{equation}
with:
\begin{equation}\label{Courrege2}
\begin{split}
 Lf(\psi) =& \frac12\sum_{i,j=1}^d\sum_{m,n\in\{R,I\}}b^{i,m}_{j,n}(\psi)\frac{\partial^2}{\partial\psi_{i,m}\partial\psi_{j,n}}f(\psi) \\
&+ \sum_{i=1}^d\sum_{m\in\{R,I\}}a_{i,m}(\psi)\frac{\partial}{\partial\psi_{i,m}}f(\psi)
+ c(\psi)f(\psi) \\
&+ \int\nu(\psi,\text d\phi) \left[ f(\psi+\phi)-f(\psi)\phantom{\sum_{i=1}^d} \right. \\
&\left.-\left(\sum_{i=1}^d\sum_{m\in\{R,I\}}\phi_{i,m}\frac{\partial}{\partial\psi_{i,m}}f(\psi)\right)1_{[0,1]}(\|\phi\|)\right] 
\end{split}
\end{equation}
\cite{Kolokoltsov, Gardiner}. Here $R$ and $I$ denote real and imaginary parts, $\psi_{i,m}$ real and imaginary parts of the coordinates of $\psi$ with respect to $e_1, e_2, \dots$, $(b)$ a $\psi$-dependent, 
real ($2d\times2d$), symmetric and nonnegative definite matrix and $\nu$ a stochastic kernel. (This formula is usually formulated for $R^d$-valued processes. It cannot be transferred to $C^d$ by reading all products as complex ones, 
but one has to consider $C^d$ as a real vector space with double dimension and do all calculations with real numbers.)
Substituting $f_{k,l}(\psi):=\left(\langle e_k,\psi\rangle\langle\psi,e_l\rangle\right)_R$ with $k,l\ge2$ into Eqs.~\eqref{Courrege1} and~\eqref{Courrege2}, one gets for $t=0$ and $\psi=\varphi$:
\begin{equation}\label{senkrecht2a}
\begin{split}
 &\left.\frac{\partial}{\partial t}\right|_{t=0,\psi=\varphi}f_{k,l} = \left.\frac{\text d}{\text dt}\right|_{t=0}\langle e_k,\rho_te_l\rangle_R \\
=& \; b^{k,R}_{l,R}(\varphi) + b^{k,I}_{l,I}(\varphi) + \int(\langle e_k,\phi\rangle\langle\phi,e_l\rangle)_R\nu(\varphi,\text d\phi)
\end{split}
\end{equation}
(all terms in Eq.~\eqref{Courrege2} except the ones involving second derivatives and the integral of $f(\psi+\phi)$ contain factors like $\langle e_k,\varphi\rangle$ with $k\ge2$ and therefore vanish).
Similarly, for $g_{k,l}(\psi):=\left(\langle e_k,\psi\rangle\langle\psi,e_l\rangle\right)_I$ one gets:
\begin{equation}\label{senkrecht2b}
\begin{split}
 &\left.\frac{\partial}{\partial t}\right|_{t=0,\psi=\varphi}g_{k,l} = \left.\frac{\text d}{\text dt}\right|_{t=0}\langle e_k,\rho_te_l\rangle_I \\
=& \; b^{k,I}_{l,R}(\varphi) - b^{k,R}_{l,I}(\varphi) + \int(\langle e_k,\phi\rangle\langle\phi,e_l\rangle)_I\nu(\varphi,\text d\phi) \,.
\end{split}
\end{equation}
Now we choose a symmetric real $2d\times2d$ matrix $(\sigma^{k,o}_{l,p})$ such that $b^{k,o}_{l,p}(\varphi) = \sum_{i,m}\sigma^{k,o}_{i,m}\sigma^{l,p}_{i,m}$.
Reading it as a $d\times2d$ complex matrix representing an operator $\sigma: \mathbb C^{2d}\to\mathbb C^d$, Eqs.~\eqref{senkrecht2a} and~ \eqref{senkrecht2b} can be written more concisely as:
\begin{eqnarray}
 \left.\frac{\text d}{\text dt}\right|_{t=0}\langle e_k,\rho_te_l\rangle & = & \langle e_k,\sigma\sigma^\dagger e_l\rangle \nonumber \\
& + & \langle e_k, \left(\int |\phi\rangle\langle\phi|\nu(\varphi,\text d\phi)\right)e_l\rangle \,. \nonumber
\end{eqnarray}
Together with Eq.~\eqref{senkrecht1}, one gets:
\begin{equation*}
 \sum_k c_k|L_k\varphi\rangle_\bot\langle L_k\varphi|_\bot = (\sigma\sigma^\dagger)_\bot + \left(\int |\phi\rangle\langle\phi|\nu(\varphi,\text d\phi)\right)_\bot \,.
\end{equation*}
$\sigma\sigma^\dagger$ and the integral are nonnegative, consequently also their orthogonal parts, so also the left-hand side has to be nonnegative.
Since $\varphi,L_1\varphi,\dots$ are linearly independent, the same is true for
$(L_1\varphi)_\bot, (L_2\varphi)_\bot, \dots$, therefore all $c_k$ have to be nonnegative. This is complete positivity, and Eq.~\eqref{lindblad+-} is a Lindblad equation.

Concerning the proof of the second theorem, the result has been also obtained in \cite{WisemanDiosi} and the proof in our setting is presented in  the appendix.

\noindent \emph{Fokker-Planck equation.}
It is interesting to understand in which way the probabilities for the wave functions develop on Hilbert space which then allow to classify the possible diffusions
in more detail. For example 
we note that some diffusions  have the same diffusive term in the Fokker Planck equation governing the probability distribution on Hilbert space: 
The Fokker-Planck equation for the probability
density $p$ with respect to the ``Hilbert space volume element" $\mathcal D\psi_R\mathcal D\psi_I$ (R and I denoting real and imaginary part) reads:
\begin{eqnarray}
\lefteqn{ \frac{\partial p_t(\psi)}{\partial t} = - \sum_{m\in\{R,I\}}\int \left[\frac{\partial \left(A(\psi)_mp(\psi)\right)}{\partial \psi(x)_m}\right](x)\text dx} &&\\
&+&\frac12\sum_{l,m\in\{R,I\}}\int\int\left[\frac{\partial^2 \left(D(\psi)p(\psi)\right)}{\partial \psi(x)_l\partial \psi(y)_m}\right](x,y)_{l,m}\text dx\text dy\,\,, \nonumber
\end{eqnarray}
if the diffusion matrix $D$ is defined via its real entries:
\begin{equation}
 [D(\psi)](x,y)_{l,m} = \sum_{k=1}^n [B_k(\psi)](x)_l[B_k(\psi)](y)_{m} \,.
\end{equation}
More precisely, choosing an orthonormal basis of the Hilbert space
and introducing the corresponding complex coordinates $x_1=(x_{1,R}, x_{1,I}), x_2, \dots$, one can write:
\begin{eqnarray}
\frac{\partial p_t(x)}{\partial t} & = & -\sum_{\stackrel{k=1,2,\dots}{m\in\{R,I\}}} \frac{\partial}{\partial x_{k,m}}\left([A(x)]_{k,m}p(x)\right) \\
& & +\frac12\sum_{\stackrel{i,j=1,2,\dots}{m,n\in\{R,I\}}}\frac{\partial^2}{\partial x_{i,m}\partial x_{j,n}}\left([D(x)]^{i,m}_{j,n}p(x)\right) \nonumber
\end{eqnarray}
 where the diffusion matrix $D$ has entries:
\begin{equation}
 [D(x)]^{i,m}_{j,n} = \sum_{k=1}^n [B_k(x)]_{i,m}[B_k(x)]_{j,n} \,.
\end{equation}
Thus, within this calculus, one has to consider the Hilbert space as a real one with double dimension. In particular, the complex matrix $\sum_k|B_k\rangle\langle B_k|$ can be computed from $D$, but not
vice versa - the distribution of $\psi_t$ is not fixed by $\sum_k|B_k\rangle\langle B_k|$.

Replacing the matrix $u$ by $ou$, where $o$ is real orthogonal, leaves not only $\sum_k|B_k\rangle\langle B_k|$, but also $D$ invariant;
in particular, for any real orthogonal $u$, the diffusion matrix is the same as for $u_{i,j}=\delta_{ij}$ (the standard collapse equation).
With complex entries, this need not be the case. The simplest example is the choice  $u_{ij}=i\delta_{ij}$ which, assuming that the $L_k$
are self-adjoint, gives $B_k(\psi)=iL_k\psi$ and $A(\psi)=-iH\psi ß \frac12\sum_kL_k^2\psi$, i.e. a linear equation reproducing Eq.~\eqref{Lindblad2} 
and generalizing Eq. (16) of the supplement.

We remark that having introduced the probability distribution $p$ allows us to write the density matrix more explicitly as the second moment of the probability distribution on the Hilbert space, namely:
\begin{equation}
\rho_t(x,y)=\int\psi(x)\overline{\psi(y)}\mathcal D[\psi]\,.
\end{equation}

\noindent {\it The Born rule.} Collapse model predict that outcomes are distributed according to the Born rule~\cite{adlernftlr2,ab1}. As first noted in~\cite{adlernftlr,adlernftlr2}, this means that this rule is the only probability rule, which is compatible with the no superluminal signalling condition . In fact, let
\begin{equation}
T[|\psi\rangle\langle\psi|] \;  = \; \sum_n p_n \, \frac{P_n |\psi\rangle\langle\psi| P_n}{|| P_n |\psi\rangle ||^2}
\end{equation}
describe the collapse, with $P_n$ projection operators. $T$ is trace preserving (the analog of norm conservation). If one also requires $T$ to be linear, then the only possible choice is to take $p_n = || P_n |\psi\rangle ||^2$, which is the Born's statistical law.


\noindent {\it Appendix. Proof of theorem 2.}
\noindent {\it Step 1.}
By using It\^o calculus~\cite{Ito}, one can show that Eq.~(4) leads to the following master equation for $\rho_t:=\mathbb E[|\psi_t\rangle\langle\psi_t|]$:
\begin{eqnarray}\label{master2}
\lefteqn{\frac{\text d\rho_t}{\text dt} =} && \\ 
& \!\!= & \!\!\mathbb E \!\left(\!|A(\psi_t)\rangle\langle\psi_t| + |\psi_t\rangle\langle A(\psi_t)| + \sum_{k=1}^N|B_k(\psi_t)\rangle\langle B_k(\psi_t)|\! \right)\!\!. \nonumber
\end{eqnarray}
Let us choose a deterministic initial value $\psi$ in Eq.~(4). Correspondingly, we set $\rho_0:=|\psi\rangle\langle\psi|$ as the initial value in Eq.~(3). Let us set both Eq.~\eqref{master2} and~(3) equal at time $t=0$. This will be sufficient, because the equations are Markovian and time-homogenous. We have: 
\begin{eqnarray}\label{20}
\lefteqn{|A(\psi)\rangle\langle\psi| + |\psi\rangle\langle A(\psi)| + \sum_{k=1}^N|B_k(\psi)\rangle\langle B_k(\psi)| =} \qquad \\
& \!\!\!\!\!\!\!\!\!\!\!\!\!\!\!\!\!\!\!\!= & \!\!\!\!\!\!\!\!\!\!\! |-iH\psi\rangle\langle\psi| + |\psi\rangle\langle-iH\psi| + \nonumber \\
& \!\!\!\!\!\!\!\!\!\!\!\!\!\!\!\!\!\!\!\!+ & \!\!\!\!\!\!\!\!\!\!\! \sum_{k=1}^n\left(|L_k\psi\rangle\langle L_k\psi|-\frac12 |L^{\dagger}_k L^{\phantom{\dagger}}_k \psi\rangle\langle\psi| -\frac12|\psi\rangle\langle L^{\dagger}_k L^{\phantom{\dagger}}_k\psi| \right) . \nonumber
\end{eqnarray}

\noindent {\it Step 2.}
Since $B_k$ and $L_k$ are the only operators appearing in the ``bra'' and ``ket'' part simultaneously, then the $B_k$ are essentially fixed by the $L_k$. To find out how, we restrict Eq.~\eqref{20} to the orthogonal complement of $\psi$. Writing the corresponding orthogonal decomposition of vectors as $v=v_{\bot}+v_{\|}$, we get:
\begin{equation}
 \sum_{k=1}^{N} |B_{k}(\psi)_{\bot}\rangle\langle B_{k}(\psi)_{\bot}| = \sum_{k=1}^n |(L_k\psi)_\bot\rangle\langle (L_k\psi)_\bot|\,.
\end{equation}
The linear independence implies $N\ge n$. Moreover,
\begin{equation}
B_k(\psi)_{\bot} \; = \; \sum_{j=1}^{N} \, u_{kj}(\psi)\, (L_j\psi)_\bot,
\end{equation}
where $L_{n+1}:=\dots:=L_N:=0$ and the functionals $u_{kj}(\psi)$ are the coefficients of a $\psi$-dependent {\it unitary} matrix~\cite{Nielsen}. Therefore we can write:
\begin{equation}
B_k(\psi) \; = \; \sum_{j=1}^{N} \, u_{kj}(\psi)\, (L_j\psi)_\bot \, - \, \tilde{\ell}_k^{(\psi)}\psi,
\end{equation}
for unspecified functionals $\tilde{\ell}_k^{(\psi)}$ of $\psi$. Equivalently, we can write:
\begin{equation}
B_k(\psi) \; = \; \sum_{j=1}^{N} \, u_{kj}(\psi)\, L_j\psi \, - \, {\ell}_k^{(\psi)}\psi \, = \, L^{(\psi)}_k \psi  \, - \, {\ell}_k^{(\psi)}\psi  \label{Bmatrix}
\end{equation}
where $L^{(\psi)}_k $ is defined as~(8), and we have defined:
\begin{equation}\label{B2}
{\ell}_k^{(\psi)}\psi \, = \,  \tilde{\ell}_k^{(\psi)}\psi + \sum_{j=1}^{N} \, u_{kj}(\psi)\, (L_j\psi)_{\|} \, .
\end{equation}

\noindent {\it Step 3.}
We now determine the operator $A$. For this,
we substitute the previous result into Eq.~\eqref{20} and we multiply with $\psi$ from the left and $v\bot\psi$ from the right. We get:
\begin{eqnarray}
 \langle A(\psi), v\rangle - \sum_{k=1}^N \ell_k^{(\psi)}\langle L_k^{(\psi)}\psi,v\rangle & = & i\langle H\psi, v\rangle -\\
&- & \frac12\sum_{k=1}^N \langle L_k^\dagger L_k\psi, v\rangle. \nonumber
\end{eqnarray}
Since $v$ is a generic vector orthogonal to $\psi$, the above equation implies that the parts of $A(\psi) -  \sum_{k=1}^{N} \ell_k^{(\psi)*}L_k^{(\psi)}\psi$ and of $-iH -\frac12\sum_{k=1}^{N} L_k^\dagger L_k \psi$ orthogonal to $\psi$ coincide, while the parallel part of $A(\psi)$ is arbitrary. Therefore, similarly to the previous step, one can write:
\begin{equation}\label{a2}
 A(\psi)= -iH\psi - \frac12 \sum_{k=1}^{N}\left( L_k^\dagger L_k\psi - 2{\ell_k^{(\psi)*}}L_k^{(\psi)}\psi \right) +G(\psi)\psi
\end{equation}
with arbitrary complex functional $G$. 

\noindent {\it Step 4.}
Substituting Eqs.~\eqref{Bmatrix} and~\eqref{a2} into~\eqref{20} and multiplying from both sides with $\psi$ gives:
\begin{equation} \label{eq:fgfggfdfc}
 G(\psi)=-\frac12\sum_k |\ell_k^{(\psi)}|^2 +ig(\psi)
\end{equation}
with real functional $g$. 

\noindent {\it Step 5.} Now we apply the condition of norm preservation for the diffusion process given by Eq.~(4). The stochastic differential equation for the square norm of $\psi_t$ is:
\begin{eqnarray}\label{norm}
 \text d\|\psi_t\|^2 & \!\!\! = & \!\!\!\! \left( \!\! \langle\psi, A(\psi)\rangle + \langle A(\psi), \psi\rangle + \sum_{k=1}^N\langle B_k(\psi), B_k(\psi)\rangle
 \!\! \right) \! \text dt  \nonumber \\
& \!\!\! + & \!\! \sum_{k=1}^N \left(\langle\psi, B_k(\psi)\rangle + \langle B_k(\psi), \psi\rangle \right)\text dW_{k,t} \,
\end{eqnarray}
which should be zero, for the norm to be conserved. By virtue of \eqref{Bmatrix},\eqref{B2},\eqref{a2},\eqref{eq:fgfggfdfc}, 
the drift term is zero. Hence also the diffusion term must be zero. Multiplying and taking expectation and recalling that it is sufficient to consider $t=0$ we obtain
\begin{eqnarray}
\lefteqn{\sum_{k=1}^{N} \left(\langle\psi, B_k(\psi)\rangle + \langle B_k(\psi), \psi\rangle \right) \mathbb{E}\left(\text dW_{k,t} \text dW_{j,t}\right)= }  \qquad\qquad\quad  \nonumber \\
& &\left(\langle\psi, B_j(\psi)\rangle + \langle B_j(\psi), \psi\rangle \right) \text dt=0\,\qquad
\end{eqnarray}
hence, in view of \eqref{Bmatrix} and \eqref{B2},
\begin{equation}
\langle \psi,L_k^{(\psi)}\psi\rangle + \langle L_k^{(\psi)}\psi,\psi\rangle - 2\, \text{Re} [\ell_k^{(\psi)}] = 0, \quad  k=1,\ldots {N}
\end{equation} 
or:
\begin{equation}\label{F}
 \ell_k^{(\psi)} = \frac12\langle \psi, (L^{(\psi)\dagger}_k + L_k^{(\psi)})\psi\rangle - ih_k(\psi) \,,
\end{equation}
for arbitrary real-valued functionals $h_k(\psi)$.  Eq.~\eqref{Bmatrix}, with  $ {\ell}_k^{(\psi)}$ given by Eq.~\eqref{F}, identifies the operators $B_k(\psi)$. Similarly, Eq.~\eqref{a2}, with $G(\psi)$ given by Eq.~\eqref{eq:fgfggfdfc}, identifies the operator $A(\psi)$. So far, the unspecified quantities are the functionals $h_k(\psi)$ and $g(\psi)$, as well as the coefficients $u_{kj}(\psi)$.

It is easy to see that the functionals $g$ and $h_k$ are unimportant global phase factors. In fact, $\varphi_t:=e^{-i\int_0^tg(s)\text ds}\psi_t$ satisfies Eq.~\eqref{master2} with $g=0$. In a similar way, $\phi_t:=e^{- i\sum_{k=1}^n \int_0^t  h_k(s) dW_{k,s}}\varphi_t$ satisfies
Eq.~\eqref{master2} with $g=h_k=0$, as a simple application of  It\^o's formula shows. Therefore the only degree of freedom is encoded in the the unitary matrix with coefficients $u_{kj}(\psi)$.

By straightforward use of It\^o calculus, the operators $A$ and $B_k$ we have found, indeed reproduce the Lindblad equation~(3). Therefore also the `if' part is proven. This concludes the proof of the statement.

To understand the meaning of the unitary freedom, let us consider the simpler case of only one self-adjoint Lindblad operator $L$. In such a case, the unitary matrix reduces to a complex phase factor $e^{if(\psi)}$. Eq.~(4), with the previous specifications for $A$ and $B_k$, becomes:
\begin{eqnarray}\label{collapse3}
\text d\psi_t & = &  -i H \psi_t \text dt + \left[ e^{if(\psi_t)}L \psi_t  \right. \nonumber \\
& & - \langle \psi_t, L \psi_t \rangle \psi_t \cos (f(\psi_t)) \Big] \text dW_t\, \nonumber \\
& & - \frac12 \left[ L^2 \psi_t - 2 e^{if(\psi_t)} \langle \psi_t, L \psi_t \rangle L \psi_t \cos (f(\psi_t))  \right.\nonumber \\
& & + |\langle \psi_t, L \psi_t \rangle|^2 \psi_t \cos^2 (f(\psi_t)) \Big] \text dt,
\end{eqnarray}
When $f=0$, one recovers the standard collapse equation~\cite{physrept,rmp}. For $f = \pi/2$ instead, all non-linear terms vanish: one obtains a linear and random evolution, which, written in the Stratonovich formalism---which is closer to the physicists' usual formalism---becomes:
\begin{equation} \label{eq:str}
\frac{\text d\psi_t}{\text dt} = -i\left[ H  - L w_t \right] \psi_t,
\end{equation}
with $w_t = \text{d} W_t/ \text{dt}$. This is a standard Schr\"odinger with a random potential, which does not give rise to the collapse of the wave function (since it is linear), but nevertheless originates the same Lindblad equation for the density matrix. This is an entirely expected result, and the previous literature is aware of such a fact, i.e. that there are infinite stochastic unravellings of the same master equation~\cite{int1,int2}. 

One way to understand the meaning of $f$ is to say that it measures the ``anti-hermitian'' part of the coupling of the noise with the wave function, which is the one responsible for the collapse. The maximum such coupling is obtained precisely for $f=0$. It is in fact, let us neglect the Hamiltonian $H$ and let us compute the equation for the variance $V_t :=  \langle \psi_t, L^2 \psi_t \rangle - \langle \psi_t, L \psi_t \rangle^2$, which is a measure of how the stavector is delocalized over the eigenstates of $L$. One obtains~\cite{ah}:
\begin{eqnarray}
\text d V_t & = &  -4 \cos^2 (f(\psi_t)) V^2_t  \text dt \nonumber \\
& & + 2 \cos (f(\psi_t))  \langle \psi_t, (L - \langle \psi_t, L \psi_t \rangle)^3 \psi_t \rangle \text dW_t \, . \qquad
\end{eqnarray}
The drift term (which proportional to $- V^2_t$) drives $V_t$ to zero for large times and for almost any realization of the noise, as expected from a collapse equation. The rate is proportional to $\cos^2 (f(\psi_t))$ and of course is maximal for $f = 0$. This is the standard choice for collapse models.

\emph{Acknowledgements.}
A.B. acknowledges partial financial support from the EU project NANOQUESTFIT, the John Templeton Foundation project
`Experimental and theoretical exploration of fundamental limits of quantum mechanics', INFN and COST (MP1006). He wishes to thank S.L. Adler, F. Benatti and S. Weinberg for stimulating discussions. D.D. and G.H. acknowledge helpful discussions with M. Kolb.
The authors are grateful to L. Diosi for pointing out reference~\cite{WisemanDiosi}, and for helpful discussions.


\begin{thebibliography}{99}
\bibitem{comment} There is an increasing interest in non-Markovian dynamics, also in the context of collapse models~\cite{nm}. Nevertheless, Markovian dynamics still remain the cornerstone for complex systems, and all fundamental equations of Physics are Markovian. This justifies restricting the analysis  to Markovian evolutions.
\bibitem{op} More precisely, and with reference to the calculations in~\cite{adlernftlr2}, linearity of the operators is not used up to 
Eq. (6.13a)---just standard It\^o calculus is used. However, in going from (6.13a) to (6.13c) linearity is used in pulling 
the stochastic expectation onto $\hat{\rho}$. Moreover, $H_{\text{\tiny eff}}$, $M^R$ and $M^I$ are assumed to be self-adjoint, in calculating 
the time derivatives of the norm and density matrix.
\bibitem{equiv} Two ensembles are said to be {\it equivalent} if they generate the same density matrix: $\sum_k p_k|\psi_{k} \rangle\langle \psi_{k}| = \sum_k q_k|\phi_{k} \rangle\langle \phi_{k}|$, where $\{\psi_k\}, \{\phi_k \}$ are the vectors belonging to the two ensembles and $p_k, q_k$ the associated probabilities, respectively.
\end{thebibliography}
\end{document}